\documentclass[prl,twocolumn,superscriptaddress,showpacs,amsmath,amssymb]{revtex4}

\usepackage{graphicx}
\usepackage{latexsym}
\usepackage{amsmath}
\usepackage{amssymb}
\usepackage{amsfonts}
\usepackage{color}
\usepackage{bm}
\usepackage{verbatim}
\usepackage{changes}

\begin{document}
\title{Long-range spin and charge accumulation in mesoscopic superconductors with Zeeman splitting}
\date{\today}

\author{M. Silaev}
 \affiliation{O.V. Lounasmaa Laboratory, P.O. Box 15100, FI-00076 Aalto University, Finland}

\affiliation{Department of Theoretical Physics, The Royal
Institute of Technology, Stockholm SE-10691, Sweden}

\author{P.~Virtanen}
\affiliation{O.V. Lounasmaa Laboratory, P.O. Box 15100, FI-00076
Aalto University, Finland}

\author{F.S.~Bergeret}

\affiliation{ Centro de F\'{i}sica de Materiales (CFM-MPC), Centro
Mixto CSIC-UPV/EHU, and Donostia International Physics Center (DIPC), Manuel de Lardizabal 5, E-20018 San
Sebasti\'{a}n, Spain}

\author{T.T.~Heikkil\"a}

 \affiliation{Department of
Physics and Nanoscience Center, University of Jyv\"askyl\"a, P.O.
Box 35 (YFL), FI-40014 University of Jyv\"askyl\"a, Finland}

 \begin{abstract}
 We describe the far from equilibrium non-local transport in a diffusive superconducting
 wire with a Zeeman splitting, taking into account the different spin
 relaxation mechanisms.
 We demonstrate that due to the Zeeman splitting an injection of a current in a superconducting wire creates a spin accumulation that can only relax via thermalization. In addition
 the Zeeman splitting also  causes a suppression of the  spin-orbit and spin-flip scattering rates.
 These two effects lead to long-range spin and charge accumulations detectable
 in the non-local signal.
 Our model explains the main qualitative features of recent experimental results in terms
 of realistic parameters and predicts a strong dependence of the
 non-local signal on the orbital depairing effect from an induced magnetic field.
 \end{abstract}

\maketitle

\begin{figure}[!htb]
 \centerline{\includegraphics[width=1.0\linewidth]{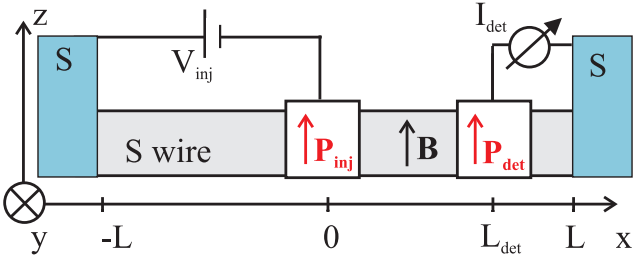}}
 \caption{\label{Fig:Sketch} (Color online) Schematic view of the
   setup for nonlocal conductance measurements. Here we assume that
   the polarizations of the magnetic contacts are collinear to the
   magnetic field,
 ${\bm P_{inj}}\parallel {\bm P_{det}}\parallel {\bm B}$. }
 \end{figure}

Hybrid ferromagnetic/superconducting (FS) structures reveal a
rich physics originating from the interplay between magnetism and
superconductivity \cite{BuzdinRMP,Bergeret2001}. While most of the
research activity has been focused on the study and detection of
proximity induced triplet superconducting correlations in an
equilibrium situation \cite{Bergeret2001,triplet}, more recent
experiments addressed the problem of spin and charge accumulation
in superconducting
wires\cite{Fukuma2011,HanleSuper,Poli,SpinInjectionNb,Aprili2013,Beckman2012,Beckman2014}.
Figure \ref{Fig:Sketch} shows  a typical experimental setup, in
which a  spin accumulation is generated by a spin-polarized
current injected from a ferromagnetic electrode. This spin
accumulation observed in the experiments can be quite large. Two
puzzling findings motivate this Letter: First, in superconductors
with a strong Zeeman splitting, the induced spin accumulation has
been detected at distances from the injector much larger than the
spin-relaxation length in the normal
state \cite{Aprili2013,Beckman2012,Beckman2014}. Second, the
non-local conductance $g_{nl}$ depends drastically  on the origin
of the Zeeman splitting. Such a splitting can be caused either by
an applied (strong) external magnetic
field \cite{Aprili2013,Beckman2012} or by the proximity of a ferromagnetic
insulator \cite{Beckman2014proximity}.

In this Letter,  we develop a microscopic model based on
 the well-established Keldysh kinetic equations for superconductors
  extended to spin-dependent phenomena, and solve this puzzle.
In particular we show that: (i) The observed long-range spin
accumulation can be understood as a thermoelectric effect for
Bogoliubov quasiparticles. The heating of a superconducting wire,
originated for example from an injected current, produces a spin
accumulation which can be detected as an electric signal by a
spin-filter detector.
 The spin accumulation created in such a way can relax only due to
the thermalization of injected quasiparticles and therefore
 the spin relaxation length is determined by inelastic
electron-phonon and electron-electron scattering that can well
exceed the usual spin diffusion length. (ii) Besides generating a
large thermoelectric effect the Zeeman splitting also
 suppresses the spin-flip and spin-orbital scattering which are
the main sources of charge imbalance relaxation in superconductors
at low temperatures \cite{PairBreakingChImb}.
 Hence the different behaviors observed for the non-local conductance $g_{nl}$ as a function of the injection voltage $V_{inj}$,
depend on the value of the  orbital depairing parameter
$\alpha_{orb}$ defined below.
For large enough values of $\alpha_{orb}$, at large applied fields
 the contribution from the  charge imbalance to the non-local conductance is suppressed
 and the $g_{nl}(V_{inj})$ dependence is almost antisymmetric with respect to $V_{inj}$ \cite{Aprili2013,Beckman2012,Beckman2014}.
 In contrast,   if the Zeeman splitting
 is caused by the proximity of
 a ferromagnetic insulator \cite{Beckman2014proximity},
  $\alpha_{orb}$ is small and the charge imbalance contribution to $g_{nl}$ becomes  important.
  In this case, we predict a  qualitative change
of the  non-local conductance as function of the injected current,
that can be experimentally proven.

We consider the nonlocal spin valve shown in
Fig.~\ref{Fig:Sketch}. A spin-polarized current is injected in the
superconducting wire from a ferromagnetic electrode with
polarization ${\bm P_{inj}}$, pointing in the direction of the
magnetization. The detector is also  a ferromagnet with a
polarization vector ${\bm P_{det}}$ and located at a distance
$L_{det}$ from the injector. Both the injector and the detector
are coupled to the wire via tunnel contacts. A magnetic field
${\bm B}$ is applied in $z$ direction.

  When ${\bm P_{inj}}\parallel {\bm B}\parallel {\bm P_{det}}$ (for
  the non-collinear case, see Ref.~\cite{silaevup14}),
 the tunnelling current at the detector is given by
   \begin{equation}\label{Eq:ZeroCurrentYGen}
  R_{det}I_{det}= \mu  +  P_{det} \mu_z
    \end{equation}
where $R_{det}$ is the detector interface resistance in the normal
state, $\mu$ is the charge imbalance and $\mu_z$ the spin
imbalance. Here we assume that the detector current is measured at
zero bias $V_{det}=0$.  The nonlocal differential conductance
measured in the experiment is $ g_{nl}= d I_{det}/d V_{inj} $.


  The charge imbalance $\mu$ and spin accumulation $\mu_z$ can be expressed in terms of the
  Keldysh quasiclassical Green function (GF) as
  $ \mu = \int_{0}^\infty  {\rm Tr}(g^K) d\varepsilon/16 $ and
$ \mu_{z} = \int_{0}^\infty  {\rm Tr}[\tau_3 \sigma_3
(g^K-g^K_{eq})] d\varepsilon/16 $.
    Here $\tau_3$ ($\sigma_3$) is the third Pauli matrix in Nambu (spin) space,
 $g^K$ is the (4$\times$4 matrix) Keldysh component of the quasiclassical GF matrix  $\check{g} = \left(%
 \begin{array}{cc}
  g^R &  g^K \\
  0 &  g^A \\
 \end{array}%
 \right)$, and $g^{R(A)}$ is the retarded (advanced) GF. We denote $g^K_{eq}=g^K$ at the equilibrium state. The matrix GF satisfies the normalization
 condition $\check g^2=1$ that allows writing the Keldysh component as
 $ g^K= g^R  \hat f - \hat f  g^A$, where  $\hat f$ is the distribution function  with a general spin structure
 $ \hat f= f_L +f_T\tau_3  +  f_{T3} \sigma_3+  f_{L3}
 \tau_3\sigma_3$\cite{SupplMat}.
 With the help of the above notations we obtain the expressions for the
 charge and spin imbalance in the superconductor (here and below, $\hbar=k_B=1$)
 \begin{eqnarray} \label{Eq:ChPot0}
   \mu = \frac{1}{2}\int_{0}^{\infty} d\varepsilon ( N_+ f_T+ N_-  f_{L3}) \\\label{Eq:ChPotZ}
   \mu_z = \frac{1}{2}\int_{0}^{\infty} d\varepsilon [ N_+ f_{T3}+ N_-(f_{L}-n_0)],
   \end{eqnarray}
   where $N_+$ is the total density of states (DOS), $N_-$
    is the DOS difference between the spin subbands,  and $n_0(\varepsilon) = \tanh(\varepsilon/2T)$.

   According to  Eq.~(\ref{Eq:ChPotZ})  there are two contributions to the spin signal. One
 is generated from the longitudinal component $f_L$. This contribution
 is only finite in the presence
   of a Zeeman splitting of the  DOS ($N_-\neq 0$). The second contribution is described
   by the first term in the integrand of Eq.~(\ref{Eq:ChPotZ}) and it is finite even in the absence of an exchange field.
   While this latter contribution has been analyzed in
   Ref.~\cite{Beckman2012}, we show below that in several cases it is the longitudinal contribution that dominates
   the spin signal due to its long-range character.

  In order to obtain the kinetic equations in a diffusive spin-polarized superconductor we start from the general Usadel equation \cite{Bergeret2001}
  \begin{equation}\label{Eq:Usadel1}
 D\nabla\cdot(\check{g}\nabla\check{g})+ [\check\Lambda - \check\Sigma_{so} - \check\Sigma_{sf} - \check\Sigma_{orb}, \check{g}] =0.
 \end{equation}
   Here $D$ is the diffusion constant, $\check\Lambda = i\varepsilon \tau_3-i({\bm{ h\cdot S})}\tau_3 - \check{\Delta}$, $\varepsilon$ is the energy,
   $ \check{\Delta}=\Delta\tau_1$ the spatially homogeneous order parameter in the wire,
  ${\bm h }$ is the Zeeman field,
  and ${\bm S}= (\sigma_1,\sigma_2,\sigma_3)$ the vector of Pauli matrices in spin space. The last three terms in
   Eq.~\eqref{Eq:Usadel1},  $\check\Sigma_{so} = \tau_{so}^{-1} ({\bm {S}}\cdot\check{g} {\bm
   {S}})$,   $\check\Sigma_{sf} = \tau_{sf}^{-1} ({\bm {S}}\cdot\tau_3\check{g} \tau_3 {\bm
   {S}})$ and $\check\Sigma_{orb} = \tau_{orb}^{-1} \tau_3\check{g} \tau_3$
 describe spin and charge imbalance relaxation due to the
 spin-orbit scattering, exchange interaction with magnetic
 impurities and orbital magnetic depairing, characterized by the relaxation times $\tau_{so}$,
 $\tau_{sf}$ and $\tau_{orb}$, respectively.

  The orbital depairing rate can be written in the form $\tau^{-1}_{orb} = T_c \alpha_{orb} (h/T_c)^2$ where
  $\alpha_{orb}$ is the dimensionless  parameter measuring the relative strength of orbital and
  paramagnetic effects and $T_c$ is the critical temperature of the
  superconductor for $h=0$.   If the Zeeman field is provided by an external magnetic
   field \cite{Beckman2012, Aprili2013} ${\bm h}= \mu_B {\bm B}$ where $\mu_B$ is the Bohr magneton then
   $\alpha_{orb}=  T_c\Delta/(\mu^2_B B^2_c) $, where
   $B_c = \sqrt{12} \phi_0/ (\pi\xi W)$ is the critical field of a
 thin superconducting film of width $W$, $\xi = \sqrt{D/\Delta}$ is the superconducting
 coherence length and $\phi_0=h/(2e)$
 is the magnetic flux quantum \cite{SchmidtDepairing}.
   Assuming $\Delta=240$ $\mu$eV and diffusion constant $D=40$ cm$^2$/s \cite{Beckman2012, Beckman2014}, we obtain
   $\alpha_{orb} \approx 210 (W/\xi)^2$ where $\xi \approx
   100$ nm. This estimation yields $\alpha_{orb}=1.33$ for the film width $W=8$ nm.
  The Zeeman field can also be induced by an exchange
   field in a FS proximity system \cite{Beckman2014proximity}. In this case $\tau_{orb}$
   is not directly related to the Zeeman field and we describe this
   with $\alpha_{orb}=0$ in the numerical results below.

 We assume that the transparencies of the detector and injector interfaces are small, so that up to leading order the
 retarded and advanced GF  are the  bulk ones determined by  the nonlinear
 equation $ [\Lambda^R-\Sigma^R_{sf}-\Sigma^R_{so}-\Sigma^R_{orb}, g^R]=0$.
In the presence of an exchange field ${\bm h} = h{\bm z}$, the
spectral functions read
 $ g^R = g_{01}\tau_1+ g_{31} \sigma_3\tau_1 + g_{03} \tau_3 + g_{33}
 \sigma_3\tau_3$
 and $ g^A=-\tau_3 g^{R\dag}\tau_3$.
 While the terms diagonal in Nambu space ($\tau_3$) correspond to the normal GFs,
 the  $g_{01},g_{31}$ describe the singlet and zero-spin triplet
 anomalous components \cite{Bergeret2001}. From these GFs we get $N_+={\rm Re} g_{03}$, $N_-={\rm Re}
 g_{33}$ in Eqs.~(\ref{Eq:ChPot0},\ref{Eq:ChPotZ}).

From Eq.~(\ref{Eq:Usadel1}) we obtain two decoupled
sets of kinetic equations
 complemented by boundary conditions
(BC) at the spin-polarized injector interface $z=0$. We use the BC
of Ref.~\cite{bergeret12} that generalizes the Kupriyanov-Lukichev
\cite{KL} one to the case of spin-dependent barrier transmission.

We start analyzing the set of equations that couple the components $f_T$ and $f_{L3}$. These determine
 the charge ${\bm {j_c}}$ and spin-heat ${\bm {j_{se}}}$ currents according to:
  \begin{eqnarray}
  {\bm {j_c}}=  {\mathcal{D}}_{T}\nabla f_{T}+{\mathcal{D}}_{L3}\nabla f_{L3} \\
  {\bm {j_{se}}}={\mathcal{D}}_{T}\nabla f_{L3}+{\mathcal{D}}_{L3}\nabla f_{T},
  \end{eqnarray}
  where the diffusion coefficients are
  \begin{eqnarray}
  {\mathcal{D}}_{T} &=& D\left( 1+|g_{01}|^2 + |g_{03}|^2 + |g_{31}|^2 + |g_{33}|^2 \right)\\
  {\mathcal{D}}_{L3} &=& 2D{\rm Re} \left(g_{03}g_{33}^* + g_{01}g_{31}^* \right)  .
  \end{eqnarray}
  These currents satisfy a pair of coupled diffusion equations
  \begin{eqnarray}\label{Eq:fTKinC}
  {\bm {\nabla\cdot j_c}} &=&
  R_{T} f_{T}  + R_{L3} f_{L3}  \\
  \label{Eq:fTKinCS}
  {\bm {\nabla\cdot j_{se}}}
  &=&  (R_{T} + S_{L3}) f_{L3}  + R_{L3} f_{T},
  \end{eqnarray}
    supplemented by the BC at the injector electrode $x=0$:
      \begin{eqnarray} \label{Eq:fTBCC}\nonumber
  {\bm {j_c}} = \frac{\kappa_I}{D} \{ N_+(f_T-n_-) + N_- [ P_I (n_0 - n_+) +f_{L3} ]  \} \\
  \label{Eq:fTBCS}\nonumber
  {\bm {j_{se}}} = \frac{\kappa_I}{D} \{ N_-(f_T-n_-) + N_+ [ P_I (n_0 - n_+) +f_{L3} ]  \},
  \end{eqnarray}
  where $n_{\pm}  = [ n_0 (\varepsilon + V_{inj}) \pm   n_0  (\varepsilon - V_{inj})
  ]/2$. The coefficients in Eqs.~ (\ref{Eq:fTKinC},\ref{Eq:fTKinCS}) are given by
  $ R_{T}=4\Delta{\rm Re} g_{01}$,  $R_{L3}=4\Delta{\rm Re}
  g_{31}$, and
  $$
   S_{L3} = \frac{16}{\tau_\Sigma} \{({\rm Re}g_{03})^2 -({\rm Re}g_{33})^2 +
  \beta\left[ ({\rm Re}g_{31})^2 -({\rm Re}g_{01})^2\right] \}.
  $$
  Here $\tau^{-1}_\Sigma = \tau^{-1}_{so} +\tau^{-1}_{sf}$ and
 the parameter $\beta= (\tau_{so}-\tau_{sf})/(\tau_{so}+\tau_{sf})$ characterizes
  the relative strength of spin-orbit and spin-flip
  scattering. For example, in {\rm Al} wires used in
  the spin-transport experiments, the typical spin relaxation time
  is $\tau_\Sigma \approx 800$ ps $\approx 40/T_{c}$
  where $T_{c} \approx 1.6$ K.
  For the spin accumulation experiments
  \cite{Beckman2012,Beckman2014,Beckman2014proximity} the value of
  $\beta$ can be inferred from the magnetic depairing parameter $\zeta
  =8 \cdot10^{-4}$ in the absence of a magnetic field. It is proportional to the
  spin-flip scattering rate $\zeta = 3(1+\beta)/(2\tau_\Sigma T_{c})
  $ which yields $\beta\approx -0.98$. We use $\beta=-0.9$ to
  obtain the qualitative effects.

\begin{figure}[!htb]
 \centerline{\includegraphics[width=1.0\linewidth]{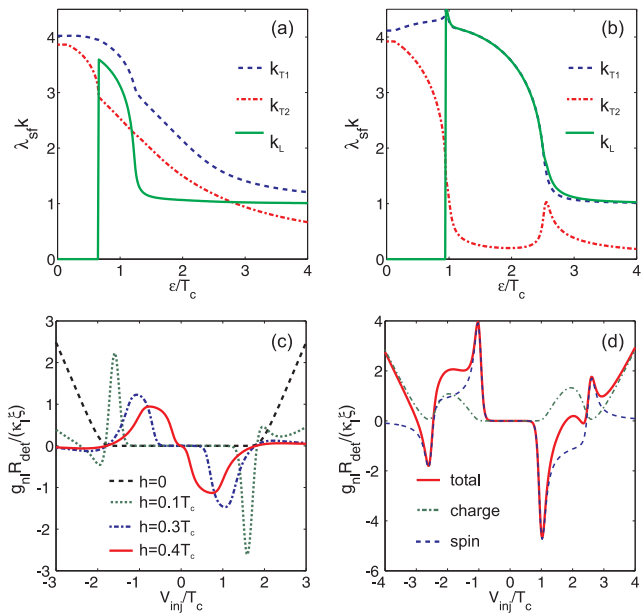}}
 \caption{\label{Fig:Scales}(Color online)
 (a,b) Energy dependence of the inverse length scales $k_{T1,2}$ and
 $k_L$ for: (a) $\alpha_{orb}=1.33$ and
 $h=0.3 T_c$;  (b)  $\alpha_{orb}=0$, and
 $h=0.8 T_c$ (here $\lambda_{sf} = \sqrt{D\tau_\Sigma/8}$ is the normal state spin relaxation length).
Panels (c) and (d) show the nonlocal conductance as a function of the
injecting voltage,
 $g_{nl}(V_{inj})$ for: (c)  $\alpha_{orb}=1.33$ and
$h=0;\; 0.1;\;0.3;\; 0.4 T_c $.
 (d) $\alpha_{orb}=0$ and $h=\;0.8T_c$. Red solid line shows the total signal,
 blue dashed and green dash-dotted lines show the spin and charge
 imbalance contributions respectively.
The parameters common to all panels are $T=0.05\; T_{c}$,
 $\beta=-0.9$, $\tau_\Sigma T_c =40$,  $P_{inj}=-P_{det}=0.1$,
 $\kappa_I\xi=0.02$,
the inelastic relaxation length $L=20\lambda_{sf}$ and
$L_{det}=5\lambda_{sf}$.  }
 \end{figure}

  The solution of the system (\ref{Eq:fTKinC},\ref{Eq:fTKinCS})
  is given by the superposition of two exponentially decaying functions $e^{-k_{T1,2}x}$ with
  amplitudes determined by the BC.
  The energy dependencies of
  $k_{T1,2}$ are shown in Fig.~\ref{Fig:Scales}a,b for the cases
  of strong and  weak orbital depairing.
  The charge and spin-heat imbalance relaxation
  is non-zero above the gap $k_{T1,2}\neq 0$ due to the magnetic pair
   breaking effects \cite{ShcmidSchoen1975, PairBreakingChImb}.

 The  other set of equations are for the components $f_L,f_{T3}$ that determine the energy and pure spin currents
  \begin{eqnarray}
   {\bm {j_e}}=  {\mathcal{D}}_{L}\nabla f_{L}+{\mathcal{D}}_{T3}\nabla f_{T3} \\
   {\bm {j_{s}}}={\mathcal{D}}_{L}\nabla f_{T3}+{\mathcal{D}}_{T3}\nabla f_{L}
   \end{eqnarray}
   satisfying  the diffusion equations
    \begin{eqnarray}\label{Eq:fLKinE}
    {\bm {\nabla\cdot j_e}}  &=&  0\\
    \label{Eq:fLKinS}
     {\bm {\nabla\cdot j_s}} &=& S_{T3} f_{T3},
 \end{eqnarray}
 where the diffusion coefficients are
 \begin{eqnarray}
    {\mathcal{D}}_{L} &=& D\left(1-|g_{01}|^2 + |g_{03}|^2 - |g_{31}|^2 + |g_{33}|^2 \right) \\
   {\mathcal{D}}_{T3} &=& 2D{\rm Re} \left(g_{03}g_{33}^* - g_{01}g_{31}^*\right)
   \end{eqnarray}
 and  $$ S_{T3} = \frac{16}{\tau_\Sigma} \{ ({\rm Re}g_{03})^2  - ({\rm Re}g_{33})^2
 + \beta \left[ ({\rm Im}g_{01})^2 - ({\rm Im}g_{31})^2 \right] \} . $$
The  boundary conditions at $x=0$ are
  \begin{eqnarray}\label{Eq:fLBCE}
 {\bm {j_e}} = \frac{\kappa_I}{D} [ N_+(f_L-n_+) + N_- (f_{T3} -P_I n_- ) ] \\
  \label{Eq:fLBCS}
 {\bm {j_s}} = \frac{\kappa_I}{D} [ N_-(f_L-n_+) + N_+ (f_{T3} -P_I n_- ) ],
  \end{eqnarray}
  where $n_\pm=n_\pm(V_{inj})$.

 The solution of the system (\ref{Eq:fLKinE},\ref{Eq:fLKinS}) is given by a superposition of two qualitatively different
 terms
   \begin{equation}\label{Eq:SolFL}
   \left( f_L\atop f_{T3} \right) =
   A\left( 1 \atop - {\mathcal{D}}_L/{\mathcal{D}}_{T3} \right)
   e^{-k_Lx}+
   \left( \alpha (x-L) \atop 0 \right),
   \end{equation}
   where  the amplitudes $(\alpha, A)$ can be found from the BC
  (\ref{Eq:fLBCE},\ref{Eq:fLBCS}).
  The first term in (\ref{Eq:SolFL})
 describes a decay of the (spectral) spin imbalance with a
 characteristic length scale
 $ k_L= \sqrt{S_{T3}{\mathcal{D}}_{L}/ ( {\mathcal{D}}_{L}^2-{\mathcal{D}}_{T3}^2)}$.
 The second term in
(\ref{Eq:SolFL}) describes the rise of quasiparticle temperature
 generated by the applied voltage $V_{inj}$. This decays only via
 inelastic scattering disregarded in the above equations, but
 discussed in more detail below.

 We now calculate the non-local conductance from
  Eq.~(\ref{Eq:ZeroCurrentYGen}) and the solutions of
  kinetic equations (\ref{Eq:fTKinC},\ref{Eq:fTKinCS},\ref{Eq:fLKinE},\ref{Eq:fLKinS}).
   First, we assume a strong orbital depairing
 $\alpha_{orb}=1.33$.
 Figure (\ref{Fig:Scales})c shows the non-local conductance
 and  describes several features observed in recent experiments
\cite{Aprili2013,Beckman2012,Beckman2014}
 that we discuss below.

 In the absence of a Zeeman field, $N_-=0$,
and therefore only the first terms in the r.h.s of Eqs. (\ref{Eq:ChPot0}-\ref{Eq:ChPotZ}) contributes to $g_{nl}$.
For $h=0$ the contribution stemming from the spin accumulation
is finite only if $P_{inj}\neq0$, which is the condition to obtain a finite $f_{T3}$. However, this function
decays over the short spin diffusion length
and therefore is negligibly small at the distances $L_{det}=
3\lambda_{sf}$ from the injector. Thus, the
detected signal in this case is mostly determined
 by the charge imbalance, Eq.~(\ref{Eq:ChPot0}).
 This explains the symmetry with respect to
the injecting voltage: $g_{nl} (V_{inj})= g_{nl} (- V_{inj})$.
The charge imbalance
contribution to  $g_{nl}$ grows
monotonically when $|V_{inj}|>\Delta_g$. This behavior is
determined by the increase of the charge relaxation scale at large
energies $k^{-1}_{T2}\rightarrow\infty$ shown in
Fig.~\ref{Fig:Scales}a.

 In the presence of a magnetic field, on the one hand, the
  charge relaxation is strongly enhanced due to the orbital depairing effect.
  This explains a strong suppression of the charge imbalance background signal by increased $h$ in
 Fig.~\ref{Fig:Scales}c. On the other hand the spin imbalance
 contribution stemming from the second term in the r.h.s of Eq.
 (\ref{Eq:ChPotZ}) is large. As shown above, this term describes
 the heat injection in the presence of a finite Zeeman field $h$
 and has a long-range behavior. This contribution leads to the
 large  peaks in $g_{nl} (V_{inj})$  shown in Fig.~\ref{Fig:Scales}c.
 In contrast to the linear thermoelectric
effect \cite{ThermoelectriEschrig,ThermoelectriOsaeta}, which is
exponentially small for temperatures well below the energy gap
$T\ll\Delta$,  a non-linear heating produced by quasiparticles
injected at voltages exceeding the energy gap explains the large
electric signal observed in the experiments
\cite{Aprili2013,Beckman2012,Beckman2014}.
 Notice that the peaks do not have
 exactly the same form so that $g_{nl}(V_{inj})\neq
-g_{nl}(-V_{inj})$.  The small deviation from the antisymmetric case is due to
 the small but finite injector polarization
$P_{inj}$,  as well as to  the presence of an
admixture of the charge imbalance signal.

Next let us consider the case of no orbital depairing
$\alpha_{orb}=0$. This may correspond to the case of a Zeeman
field caused by the proximity of a ferromagnetic insulator
\cite{Beckman2014proximity}. Figure \ref{Fig:Scales}d shows
 a clearly different behavior for $g_{nl} (V_{inj})$ with respect to the large $\alpha_{orb}$ case,
and can be observed in the experiments \cite{Beckman2014proximity}.
Now the asymmetry of the $g_{nl} (V_{inj})$ curve is much more
pronounced and the peaks are much broader. This occurs due to a
significant admixture of the long-range charge imbalance
contribution (blue dashed curves) with the spin imbalance one
(green dash-dotted curve).
  While the latter is almost perfectly antisymmetric the former
 produces symmetric peaks of $g_{nl}(V_{inj})$ at voltages  within the interval
 $\Delta-h <V_{inj} <\Delta+h$. These peaks appear due to the
 strong suppression of charge relaxation by the  Zeeman splitting
 at the energy interval $\Delta-h <\varepsilon
<\Delta+h$, in accordance to  Fig.~\ref{Fig:Scales}b.

Our  results give a qualitative explanation of experiments with
large magnetic fields  [Fig.~\ref{Fig:Scales}c]. Including an
additional constant orbital depairing which can originate from the
stray fields of ferromagnetic contacts \cite{Aprili2013} we are
able to obtain accurate fits of the experimental curves shown in
Fig.3a of Ref.~\cite{Beckman2012} using realistic parameters.
Notice that  the decay of the component $f_L$, responsible for the
long-range spin imbalance, is only limited by inelastic relaxation,
which have not been taken into account in our kinetic
equations. The observed relaxation length $\lambda\sim 1$ $\mu$m likely cannot be explained
by electron-phonon scattering, which already in the normal state leads to a much larger value
$\lambda_{ph} = \sqrt{\tau_{ph} D}\approx 20$ ${\rm \mu m} $\cite{Beckman2012,Beckman2014}.
Electron-electron scattering on the other hand can redistribute the total energy in the electron
system and damp nonequilibrium components of the signal.
In order to obtain the observed relaxation length $\lambda_{ee}\sim 1$ $\mu$m one should
assume that the e-e scattering time is  $\tau_{ee} \sim 10^{-10} s$ which is significantly
less than the known value in bulk dirty Al \cite{KlapwijkRelTime} but can be achieved in
low-dimensional samples \cite{KlapwijkRelTime2D}.
The e-e thermalization process as well as nonuniversal
properties of the heat transport in real experimental setups
could explain the suppression of the spin imbalance relaxation by the
Zeeman field \cite{Beckman2012,Beckman2014}.

  To conclude,  we have developed a theoretical framework to study
  the transport properties of superconductors with a Zeeman
  splitting. We have demonstrated that the splitting field
  leads to a strong suppression  of the relaxation of charge and spin
  imbalances created by the injected current.
  In particular, the long-range spin
  accumulation observed in recent experiments is shown to be a
  manifestation of a non-linear thermoelectric effect and it is only limited by
  the  inelastic relaxation length which can be
  larger than the spin relaxation time in normal metals by several orders of magnitude.
  Our model  gives a qualitative explanation for a wide
  range of experiments on SF nonlocal spin valves, and predicts a strong dependence
  of the non-local conductance
   on orbital depairing, characterized by $\alpha_{\rm orb}$.
  Besides explaining the properties of
  superconductor-ferromagnet structures, our theory may be
  straightforwardly extended for the
  general description of thermoelectric effects in far from
  equilibrium situations in terms of the well-established theory of
  non-equilibrium GFs.

We thank Detlef Beckmann for discussions. The work of T.T.H was supported by the
Academy of Finland Center of Excellence program and the European
Research Council (Grant No. 240362-Heattronics). P.V. acknowledges the Academy of Finland
for financial support. The work of
F.S.B. was supported by the Spanish Ministry of Economy and
Competitiveness under Project No. FIS2011-28851-C02- 02 and the
Basque Government under UPV/EHU Project No. IT-756-13.

\begin{widetext}

\section{Supplementary Material}

We express the charge density, spin density, energy density and
spin energy density in terms of the Nambu-Keldysh Green's function
and show that in the presence of a spin-splitting field their
parametrization differs from the usual case.

We wish to characterize the state of the electron system in the
superconductor via different matrix elements of the Nambu-Keldysh
Green's function. We choose the Nambu vector to be of the form
\begin{equation}
\Psi = \begin{pmatrix} \psi_\uparrow({\mathbf r},t) &
  \psi_\downarrow({\mathbf r},t) & -\psi_\downarrow^\dagger({\mathbf r},t) &
  \psi_\uparrow^\dagger({\mathbf r},t)\end{pmatrix},
\end{equation}
where $\psi_\sigma^{(\dagger)}({\mathbf r},t)$ annihilates
(creates) an electron of spin $\sigma$ in position ${\mathbf r}$
at time $t$. The Keldysh Green's function is written in terms of
the Nambu vector as
\begin{equation}
G^K(1,1') = -i \tau_3 \langle [\Psi(1),\Psi^\dagger(1')]\rangle,
\end{equation}
where the argument $1$ refers to position and time ${\mathbf
r}_1$, $t_1$. Up to a state independent factor, the charge density
can then be written as
\begin{equation}
\rho(1)=-ie {\rm Tr}[G^K(1,1)]/4 = -ie \int \frac{d {\mathbf
p}}{(2\pi)^3} \int_{-\infty}^\infty \frac{d\epsilon}{2\pi} {\rm
Tr}[G^K(\epsilon,{\mathbf p},{\mathbf r},t)]/4,
\end{equation}
where the latter form is expressed in the Wigner representation.
This is thus the particle density averaged over spin.

The spin density in direction $\hat u_j \in \{\hat u_x,\hat u_y,
\hat
  u_z\}$ is obtained from
\begin{equation}
s_j(1)=-i {\rm Tr}[\tau_3 \sigma_j G^K]/4 = -i \int \frac{d
{\mathbf p}}{(2\pi)^3} \int_{-\infty}^\infty
\frac{d\epsilon}{2\pi} {\rm Tr}[\tau_3 \sigma_j
G^K(\epsilon,{\mathbf p},{\mathbf
  r},t)]/4,
\end{equation}
where $\tau_j (\sigma_j)$ is the $j$th Pauli spin matrix in Nambu
(spin) space. The multiplication with $\tau_3$ takes care of the
chosen order of spins in the Nambu vector. In order to
characterize the non-equilibrium spin accumulation we introduce
the difference between the total spin density and the one at the
equilibrium state
\begin{equation}
s_{ej}(1)= -i \int \frac{d {\mathbf p}}{(2\pi)^3}
\int_{-\infty}^\infty \frac{d\epsilon}{2\pi} {\rm Tr} \{ \tau_3
\sigma_j [ G^K(\epsilon,{\mathbf p},{\mathbf r},t) -
G^K_{eq}(\epsilon,{\mathbf p},{\mathbf r},t) ] \}/4.
\end{equation}
As shown in the main text of the paper the spin accumulation
 ${\bf s}_{ej}$ determines the tunnelling current at the spin-polarized detector electrode in the non-local measurement scheme.

The spin-averaged energy density involves a multiplication by the
Nambu matrix $\tau_3$ to take care of the fact that whereas
particles and holes contribute to the charge an opposite amount,
they contribute an equal amount to the excitation energy. We thus
can write in the Wigner representation the (internal) energy
density
\begin{equation}
\epsilon({\mathbf r},t)= -ie \int \frac{d {\mathbf p}}{(2\pi)^3}
\int_{-\infty}^\infty \frac{d\epsilon}{2\pi} \epsilon {\rm
Tr}[\tau_3 G^K(\epsilon,{\mathbf p},{\mathbf
  r},t)]/4.
\end{equation}
In the limit of a large bandwidth this becomes very large, so it
is convenient to describe only the excess energy density compared
to some equilibrium value,
 \begin{equation}
 \epsilon_e({\mathbf r},t)= -ie \int \frac{d {\mathbf p}}{(2\pi)^3}
 \int_{-\infty}^\infty \frac{d\epsilon}{2\pi} \epsilon {\rm
 Tr} \{ \tau_3 [G^K(\epsilon,{\mathbf p},{\mathbf r},t)-G^K_{\rm eq}(\epsilon,{\mathbf p},{\mathbf r},t) ] \}/4.
 \end{equation}
Similarly to the charge density, $\epsilon_e$ is a spin-averaged
quantity. We can thus also define the energy density difference in
the two spin ensembles by
\begin{equation}
\epsilon_j({\mathbf r},t)= -ie \int \frac{d {\mathbf p}}{(2\pi)^3}
\int_{-\infty}^\infty \frac{d\epsilon}{2\pi} \{\epsilon {\rm
Tr}[\sigma_j G^K(\epsilon,{\mathbf p},{\mathbf
  r},t)]\}/4.
\end{equation}
In the absence of a ferromagnetic transition, the contribution to
this quantity comes from the small energy region related to
voltage or temperature, and therefore there is no need to remove
the equilibrium value.

We now define the quasiclassical Keldysh Green's function in the
diffusive limit (and a spherical Fermi surface) via
\begin{equation}
g^K(1)=\frac{i}{\pi} \int \frac{d \hat p}{4\pi}\int d\xi
G^K(\epsilon,\xi,\hat{p},{\mathbf r},t),
\end{equation}
where ${\mathbf p}=p \hat p$ and $\xi=p^2/(2m)-\epsilon_F$, where
$\epsilon_F$ is the Fermi energy. Employing charge neutrality
(charge density vanishes on length scales that are large compared
to the usually microscopic screening length) and including the
non-quasiclassical corrections in the standard way
\cite{RammerBook}, the electrostatic potential (instead of the
charge density that vanishes) can be written in terms of the
quasiclassical Green's function as
\begin{equation}
\mu(\vec{r},t)=-\frac{1}{16} \int_{-\infty}^\infty d\epsilon {\rm
  Tr}[g^K(\epsilon,\vec{r},t)].
\end{equation}
In the absence of a ferromagnetic transition, the other
expressions can be then straightforwardly written in terms of the
quasiclassical Green's function by using
\begin{equation}
\int \frac{d{\mathbf p}}{(2\pi)^3} = \int d\xi N(\xi) \int
\frac{d\hat
  p}{4\pi} \approx N_0 \int d\xi \int \frac{d\hat p}{4\pi},
\end{equation}
where $N(\xi)$ is the density of states in the normal state, and
we assume $N(\xi) \approx N_0$ for excitation energies close to
the Fermi surface. We hence get
  \begin{align}
 s_{ej}(1)&=-N_0  \int_{-\infty}^\infty d\epsilon {\rm Tr} \{\tau_3 \sigma_j [
 g^K(\epsilon,{\mathbf r},t) - g^K_{eq}(\epsilon,{\mathbf r},t) ]\}/8.\\
 \epsilon_e(1)&=-N_0  \int_{-\infty}^\infty d\epsilon \epsilon {\rm Tr}\{\tau_3
 [g^K(\epsilon,{\mathbf r},t)-g_{\rm eq}^K(\epsilon,\mathbf{r},t)]\}/8.\\
 \epsilon_j(1)&=-N_0 \int_{-\infty}^\infty d\epsilon {\rm Tr}[\sigma_j
 g^K(\epsilon,{\mathbf  r},t)]/8.
 \end{align}
The full quasiclassical Keldysh Green's function
\begin{equation}
\check g = \begin{pmatrix} g^R & g^K\\0 & g^A\end{pmatrix}
\end{equation}
satisfies the normalization condition $\check g^2=1$. This allows
parameterizing $g^K=g^R f - f g^A$, where in the spin-dependent
case the distribution function $f$ is parameterized by eight
functions,
\begin{equation}
f=f_L+f_T \tau_3 + \sum_j (f_{Tj} \sigma_j + f_{Lj} \sigma_j
\tau_j).
\end{equation}
 Here the L-labelled functions denote the (spin)
energy degrees of freedom and are antisymmetric in energy with
respect to the Fermi level $\varepsilon=0$ in the superconductor.
The T-labelled functions are symmetric in energy and describe the
charge/spin imbalance.

In the main text, we only concentrate on the case of collinear
spin configurations, and thereby it is enough to choose one of the
spin directions, say $j=3$. In this case the above expressions for
the local potential, nonequilibrium spin density, energy density
and spin energy density can be written as
\begin{align}
\mu(\vec{r},t)&=-\frac{1}{2} \int_{0}^\infty d\epsilon (N_+
f_T+N_- f_{L3})\\
s_{e3}(\vec{r},t)&=-N_0 \int_0^\infty d\epsilon [N_+ f_{T3}+N_- (f_L-n_0)]\\
\epsilon_e(\vec{r},t)&=-N_0 \int_0^\infty d\epsilon \epsilon
[N_+(f_{L}-n_0)+N_-
f_{T3}]\\
\epsilon_3(\vec{r},t)&=-N_0 \int_0^\infty d\epsilon \epsilon (N_+
f_{L3}+N_- f_{T}),
\end{align}
where $n_0$ is the equilibrium distribution function describing
$G^K_{\rm eq}$, $\epsilon$ has been redefined with respect to the
chemical potential of the superconductor, and we have used the
fact that the integrands are symmetric with respect to
$\epsilon=0$. These expressions are also used to define the charge
and spin imbalances in Eqs.~(2-3) of the main text. When entering
the current (Eq.~(1) of the main text), the prefactors are
absorbed in the definition of the normal-state interface
resistance.

Note that in the absence of the exchange field, the density of
states (per spin) is electron-hole symmetric. In that case the
particle/spin/energy densities could be written directly in terms
of the individual distribution functions, instead of their
combinations, whose presence reveals the strong thermoelectric
effect.

\end{widetext}

\end{document}